\documentclass[prd,twocolumn,preprintnumbers,showpacs,nofootinbib,superscriptaddress,notitlepage,floatfix]{revtex4-1}
\usepackage{CJK}
\usepackage{amsmath}
\usepackage{subfigure}
\usepackage{enumerate}
\usepackage{amssymb}
\usepackage{amsfonts}
\usepackage{mathrsfs}
\usepackage{latexsym}
\usepackage{bm}
\usepackage{amscd}
\usepackage{physics} 
\usepackage{graphicx}
\usepackage{epsfig}
\usepackage{hhline,multirow}
\usepackage{dcolumn}
\usepackage{url}
\usepackage[colorlinks=true,linkcolor=blue,citecolor=blue,urlcolor=blue]{hyperref}
\usepackage{color}
\usepackage{xcolor}
\definecolor{red}{rgb}{1,0,0}

\usepackage{indentfirst}
\usepackage{subfigure}
\usepackage{psfrag}
\usepackage{slashed}
\usepackage{float}
\usepackage{setspace}
\usepackage{mathtools}
\allowdisplaybreaks[2]

\newcommand{\bea}{\begin{eqnarray}}
\newcommand{\eea}{\end{eqnarray}}
\newcommand{\be}{\begin{equation}}
\newcommand{\ee}{\end{equation}}

\begin{document}
\preprint{LA-UR-24-30910}
\title{Origin of hadron spin based on Lattice QCD study on the charmed hadrons}

\author{Fangcheng He}
\affiliation{Center for Nuclear Theory, Department of Physics and Astronomy, Stony Brook University, Stony Brook, New York 11794, USA}
\affiliation{Los Alamos National Laboratory, Theoretical Division T-2, Los Alamos, NM 87545, USA}
\author{Jian Liang}
\affiliation{Key Laboratory of Atomic and Subatomic Structure and Quantum Control (MOE), Guangdong Basic Research Center of Excellence for Structure and Fundamental Interactions of Matter, Institute of Quantum Matter, South China Normal University, Guangzhou 510006, China}
\affiliation{Guangdong-Hong Kong Joint Laboratory of Quantum Matter, Guangdong Provincial Key Laboratory of Nuclear Science, Southern Nuclear Science Computing Center, South China Normal University, Guangzhou 510006, China }
\author{Yi-Bo Yang}
\affiliation{CAS Key Laboratory of Theoretical Physics, Institute of Theoretical Physics, Chinese Academy of Sciences, Beijing 100190, China}
\affiliation{School of Fundamental Physics and Mathematical Sciences, Hangzhou Institute for Advanced Study, UCAS, Hangzhou 310024, China}
\affiliation{International Centre for Theoretical Physics Asia-Pacific, Beijing/Hangzhou, China}
\affiliation{School of Physical Sciences, University of Chinese Academy of Sciences,
Beijing 100049, China\\
\includegraphics[scale=0.35]{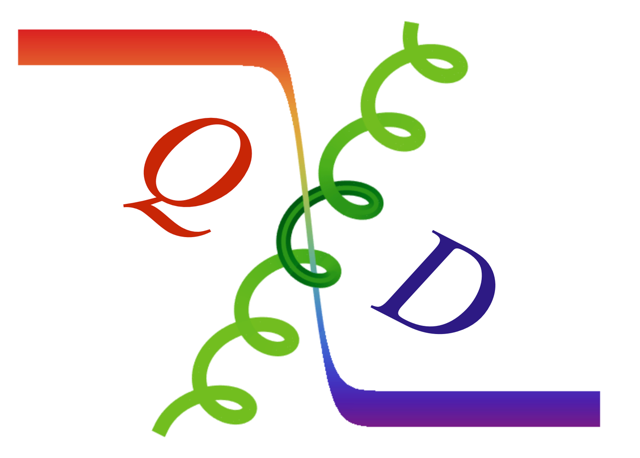}\\
($\chi$QCD Collaboration)}

\begin{abstract}
We perform the first Lattice calculation about the charmed hadron spin decomposition using overlap fermions on a 2+1 flavor RBC/UKQCD domain-wall gauge configurations at 0.083 fm with 300~MeV pion mass. It is found that the contributions of quark spin to the spin of 1S, 1P charmonia and also proton-like triple heavy quark state are comparable with the expectation of non-relativistic quark model. Such an observation provides evidence that the non-triviality of proton spin decomposition mainly arises from the relativistic effects of the light quark. Conversely, the substantial gluon angular momentum contribution in the spin $(1/2)^+$ state with triple heavy quarks at the charm quark mass, remains significant, highlighting the ongoing importance of the gluon in the realm of charmed baryon physics.
\end{abstract}

\maketitle


{\it Introduction:} Quantum chromodynamics (QCD) describes the fundamental strong interaction between quarks and gluons. Despite being formulated over 50 years ago, the understanding of how the quarks and gluons combine to form hadrons remains challenging. This difficulty arises from the non-perturbative property of QCD at low energy scale, making it impossible to perturbatively calculate how the quarks and gluon contribute to the spin of hadrons. 

For example, the naive spin sum rule suggests that two quarks have spin 1/2 while the third one has $-1/2$, resulting in a total nucleon spin of 1/2. A more careful consideration in the SU(6) quark model accounts for different flavor and spin combinations, leading to the following contributions from different flavors (see Ref.~\cite{Ma:2001ui} for further details):
\begin{align}\label{eq:delta_ud}
    \Delta u=4/3, \ \Delta d=-1/3, \ \Delta s=0.
\end{align}
However, this prediction significantly deviates from the experimental value $\Delta u-\Delta d=1.2723(23)+{\cal O}(\alpha)$ obtained from neutron weak decay, as well as from the integration of the quark polarized parton distribution function (PDF) which yields $\sum_{q=u,d,s}\Delta q\sim 25\%-30\%$~\cite{deFlorian:2008mr,deFlorian:2009vb,Blumlein:2010rn,Leader:2010rb,Ball:2013lla,Nocera:2014gqa,Ethier:2017zbq,Deur:2018roz}. 

For the missing contribution, Ji suggests a gauge-invariant and frame-independent decomposition~\cite{Ji:1996ek} of the hadron spin as 
\begin{align}\label{eq:sum_rule}
    J=\frac{1}{2}\Delta q+L_q+J_G,
\end{align}
where the total gluon angular momentum (AM) $J_g$ (and also the quark one $J_q=\frac{1}{2} \Delta q+L_q$ can be obtained from the hadron matrix element of the symmetrized energy-momentum tensor. Further decomposition of the gluon AM is possible through the Jaffe-Manohar decomposition~\cite{Jaffe:1989jz} and its extension~\cite{Chen:2008ag}, and can be accessed from Lattice QCD based on the large momentum effective theory~\cite{Ji:2013fga}. Lattice QCD calculations have confirmed the smallness of the quark spin contribution~\cite{Alexandrou:2017oeh,Green:2017keo,Liang:2018pis}, and have predicted a substantial contribution from the glue angular momentum~\cite{Yang:2016plb,Alexandrou:2017oeh,Wang:2021vqy}, which aligns with experimental indications of a non-zero glue helicity contribution~\cite{Djawotho:2013pga,Adare:2014hsq,deFlorian:2014yva, Nocera:2014gqa}.

But the origin of the tension between QCD and the quark model is far from clear. A natural guess is that the quark model is only valid when the quark mass is heavy enough, and then the smallness of the quark spin contribution will be a relativistic effect. In the case of heavy quark-antiquark pair systems, the parity and charge conjugation of a given quarkonium can be expressed as $P=(-1)^{L+1}$ and $C=(-1)^{L+S}$, where $S$ and $L$ represent the quark spin and orbital angular momentum, respectively. One can then make a simple estimation from the quark model on the quark spin and orbital angular momentum contributions to the quarkonium. 

Thus the heavy quarkonium is a non-relativistic system to verify whether the spin decomposition from Lattice QCD can be closer to that from quark model, and uncover the origin of the spin puzzle of nucleon. In this work, we will use Ji's decomposition to investigate the charmonium spin decomposition, particularly focus on the 1S($J/\Psi$) and $1P(h_c,\chi_{c1},\chi_{c2})$ channels. These channels are of particular interest due to the significant difference on the quark spin and orbital AM contributions predicted by quark model. The spin decomposition of the $(1/2)^{+}$ proton-like triple heavy quark state will also be presented to show the impact of the relativistic effect.


{\it Numerical setup:} In the context of the hadron's parton model, the quark spin contribution to the hadron is determined by the probability of the quark spin aligning with the hadron's polarization minus that in the opposite direction. Consequently, the quark spin contribution to the scalar meson is consistently zero, indicating an equal probability of the quark spin aligning in various directions. Thus, we have to calculate the matrix element of the quark spin operator ${\cal A}_z=\bar{q}\gamma_5\gamma_zq$ in the polarized hadron $H$ state,
\begin{align}
    \Delta q_{H}=&\langle H(\uparrow)|{\cal A}_z|H(\uparrow)\rangle.
\end{align}
To avoid the chiral symmetry breaking effect of ${\cal A}_z$, we use the overlap fermion~\cite{Narayanan:1994gw,Neuberger:1997fp,Liu:2002qu} which satisfies the Ginsparg-Wilson relation~\cite{Ginsparg:1981bj} $D_{ov}\gamma_5+\gamma_5D_{ov}=\frac{1}{\rho}D_{ov}\gamma_5D_{ov}$ as valence quarks to study the spin decomposition of the charmed hadron, which $\rho=1.5$ is used in this work likes the previous $\chi$QCD works. Such a discretized fermion action ensures that the effective Dirac operator $D_c=D_{ov}/(1-D_{ov}/(2\rho))$~\cite{Chiu:1998gp} has the exact chiral symmetry as that in the continuum. The calculation is carried out on the hypercubic (HYP) smeared~\cite{Hasenfratz:2001hp} 2+1 flavor domain wall fermions (DWF)~\cite{Kaplan:1992bt} ensemble 32I from the RBC/UKQCD 
collaboration, with $L^3\times T=32^3\times 64$, $a$=0.0828(3) fm and $m_{\pi}$=300 MeV~\cite{RBC:2012cbl,Mawhinney:2019cuc}. The charm quark mass is tuned by the physical $J/\psi$ mass 3.097 GeV.

Using the Feynman-Hellmann theorem inspired method~\cite{Chang:2018uxx}, we use the Coulomb gauge fixed wall source propagator $S_w(\vec{y},t_2; t_1)=\sum_{\vec{x}}S(\vec{y},t_2; \vec{x},t_1)$ and current position summed propagator with the current operator ${\cal A}$,
\bea
\tilde{S}_{c}^z(\vec{y},t_2;t_1)=\sum_{\vec{x},\tau}S(\vec{y},t_2;\vec{x},\tau)\gamma_5\gamma_zS_w(\vec{x},\tau;t_1),
\eea
to construct the two point function
\begin{align}
C_2(t_f;H)&=\langle H^p_i(t_f) H^{w\dagger}_i(0)\rangle,
\end{align}
and current position summed three point (3pt) function with the current operator ${\cal O}$
\begin{align}
\tilde{C}^q_3(t_f;{\cal O},H^{J_z})&={\cal P}^{J_z}_{ij}\langle H^p_i(t_f)\sum_{\tau}O(\tau)H_j^{w\dagger}(0)\rangle,
\end{align}
where $H_p$ ($H_w$) is the point (wall) interpolation field of the hadron state $H$. ${\cal P}^{J_z}$ is the projection matrix in both the Lorentz and Dirac space projects the hadron state to a given value of $J_z$. For the spin $1/2$ particle,  the projection can be written as ${\cal P}^{J_z}=\Gamma^+\gamma_5\gamma_z$, where $\Gamma^+=\frac{1}{2}(1+\gamma_4)$.  For the spin 1 particle, we choose ${\cal P}^{J_z}_{ij}=\epsilon^{zij}/2$. For the spin 2 state with $J_z=1$, the projection is similar to that for spin 1 case, but the hadron operator is chosen to be $T_2$ representation~\cite{Dudek:2007wv}.
Then $\Delta q_H$ for the ground state $H$ can be extract from the $t_f\rightarrow \infty$ limit of the following ratio
\begin{align}\label{eq:DeltaRq}
    \Delta R(t_f;S_q)=\frac{\tilde{C}^q_3(t_f)}{C_2(t_f)}-\frac{\tilde{C}^q_3(t_f-a)}{C_2(t_f-a)}\ _{\overrightarrow{t_f\rightarrow\infty}}\Delta q_H.
\end{align}
Details for above 3pt construction can be found in the previous $\chi$QCD works~\cite{Sun:2020pda,He:2021bof}.

Since the calculation of quark OAM can be highly non-trivial, one can calculate the gluon AM contribution first, and estimate the quark OAM through the sum rule in Eq.~(\ref{eq:sum_rule}). 

The gluon AM can be obtained using gravitation form factors (GFFs), for spin $1/2$, the GFFs can be defined using the matrix elements of energy momentum tensor (EMT) $T_{\mu\nu}^g=2\mathrm{Tr}G_{\mu}^{~\rho}G_{\rho\nu}+\frac{1}{2}g_{\mu\nu}\mathrm{Tr}G^{\rho\lambda}G_{\rho\lambda}$,
\begin{align}
\langle p'|T^g_{\mu\nu}|p\rangle=&\overline{u}(p')\Big(
A^g(q^2)\gamma^{(\mu}\bar{P}^{\nu)}+B^g(q^2)\frac{i\bar{P}^{(\mu}\sigma^{\nu)\alpha}q_\alpha}{2m_N} \nonumber\\
&\quad +C^g(q^2)\frac{q^\mu q^\nu-\eta^{\mu\nu}q^2}{m_N}\Big)u(p),
\end{align}
where $\bar{P}=(p+p')/2$, $q=p'-p$.
The gluon AM can be expressed as
\begin{align}\label{eq:JGN}
J_G^{N}=\frac{1}{2}(A^g(0)+B^g(0)).
\end{align}
On the other hand, the GFF $B$ is null in dual gravity~\cite{Mamo:2022eui}, and turns out to be small as shown in the recent Lattice calculations~\cite{Deka:2013zha,Wang:2021vqy,Hackett:2023rif} and global analysis~\cite{Guo:2023ahv}. So we neglect the contribution of $B^g(0)$ in Eq.~(\ref{eq:JGN}) for this preliminary study, and using $T^{44}$ component to calculate the matrix element in the rest frame to obtain $A^g(0)$. 

The gluon AM of spin one meson can be expressed as~\cite{Polyakov:2019lbq}
\begin{align}
    J^V_G=[J^g(q^2)+\frac12\bar{f}^g(q^2)]_{q^2\rightarrow 0},\label{eq:meson_AM}
\end{align}
where $J^g$ and $\bar{f}^g$ are the form factors which can be extracted from the hadron matrix elements of the gluon EMT using the expressions detailed in the Appendix. To extract the form factors, the 3pt correlation function can be constructed as follows
\bea
&&\tilde{C}^{g,V}_3(t_f;i,j,{\cal O},\vec{q})=\sum_{\tau=1}^{t_f-1}\int d^3\vec{y}\int d^3\vec{z}e^{i\vec{q}(\vec{z}-\vec{y})}
\nonumber\\
&&\quad \quad e^{(E_f-m)(t_f-\tau)}\langle H_i^p(t_f,\vec{y}){\cal O}(\tau,\vec{z})H_j^{w\dagger}(0)\rangle,
\eea
where $m$ is the hadron mass and $E_f=\sqrt{m^2+\vec{q}^2}$ is the energy of final state. The initial state is always chosen to be in the rest frame. 

Since the overlap factor of the final state is the same as that in the rest frame up to the discretization error, one can extract form factor $J^g(q^2)$ and $\bar{f}^g(q^2)$ using the following combination of $\Delta R^g(i,j,{\cal O},\vec{q})=\frac{\tilde{C}^{g,V}_3(t_f;i,j,{\cal O},\vec{q})}{C_2(t_f)}-\frac{\tilde{C}^{g,V}_3(t_f-a;i,j,{\cal O},\vec{q})}{C_2(t_f-a)}|_{t_f\rightarrow \infty}$, 
\begin{subequations}\label{eq:EMgluon}
\begin{align}\label{eq:EMgluon1}
&\Delta R^g(y,z,T^g_{4y},\vec{q})  \nonumber\\
&=-{\rm i}\frac{k}{4E_f}\left[(E_f+m)J^g(q^2)-(E_f-m)E^g(q^2)\right], \\
\label{eq:EMgluon2}
&\Delta R^g(z,y,T^g_{4y},\vec{q}) \nonumber\\
&={\rm i}\frac{k}{4E_f}\Big[\frac{(E_f+m)(2m^2+k^2)}{2m^2}J^g(q^2)\nonumber\\
&\quad+\Big((E_f-m)-\frac{k^2(E_f+m)}{2m^2}\Big)E^g(q^2)+2E_f\bar{f}^g(q^2)\Big], \\
\label{eq:EMgluon3}
&\Delta R^g(x,y,T^g_{xy},\vec{q})=\left[-\frac{k^2}{2}E^g(q^2)+m^2\bar{f}^g(q^2)\right]\frac{1}{2E_f},
\end{align}
\end{subequations}
where $\vec{q}=\{0,0,k\}$, $q^2=(E_f-m)^2-k^2=-k^2(1+{\cal O}(\frac{k^2}{m^2}))$. $\Delta R^g$ in Eq.~(\ref{eq:EMgluon1}) and~(\ref{eq:EMgluon2}) are imaginary due to the convention for EMT $T^{4i}$ in Euclidean space.

The form factor $\bar{f}(0)$ can be obtained in the rest frame directly from Eq.~(\ref{eq:EMgluon3}), while $J^g(0)$ can be accessed from the $q^2\rightarrow 0$ extrapolation of  $J^g(q^2)$ with non-zero $q^2$ by combining Eq.~(\ref{eq:EMgluon1}-\ref{eq:EMgluon3}). In this work, we will approximate $J^g(0)$ for the charm quark system using its value at the smallest non-zero $q^2\simeq-(2\pi)^2/L^2=-0.22$ GeV$^2$,
\begin{align}
    J^g(0)\simeq J^g(q^2)(1+\frac{q^2}{M^2_{\rm pole}})\sim J^g(q^2).
\end{align}
Given $M_{\rm pole}\sim 2m_c\sim2-3$ GeV, the systematic uncertainty introduced by this approximation is about 3\% and smaller than the statistical uncertainty.

For the $J^P=1^+$ mesons, the operator mixing can make the combination of Eq.~(\ref{eq:EMgluon1}-\ref{eq:EMgluon3}) to fail in extracting the needed form factors. Specifically, the interpolation operator for $1^{++}$ channel can mix with $0^{-+}$ state when the
momentum of final state is parallel to its polarization as the setup in Eq~(\ref{eq:EMgluon2}), and the $1^{+-}$ channel can mix with $1^{--}$ state when the momentum is perpendicular to the polarization as the setup in Eq~(\ref{eq:EMgluon1}). Again, since we are considering the charmonium with the smallest momentum allowable on this ensemble, one can assume $\bar{f}(q^2)\sim\bar{f}(0)$ and then obtain $E(q^2)$ from Eq.~(\ref{eq:EMgluon3}) in the moving frame, and eventually get $J(q^2)$ from either Eq.~(\ref{eq:EMgluon1}) (for the $1^{++}$) or Eq.~(\ref{eq:EMgluon2}) (for the $1^{+-}$). More details can be found in the Appendix.


\begin{figure}[htbp]
	\centering
    \includegraphics[width=0.38\textwidth]{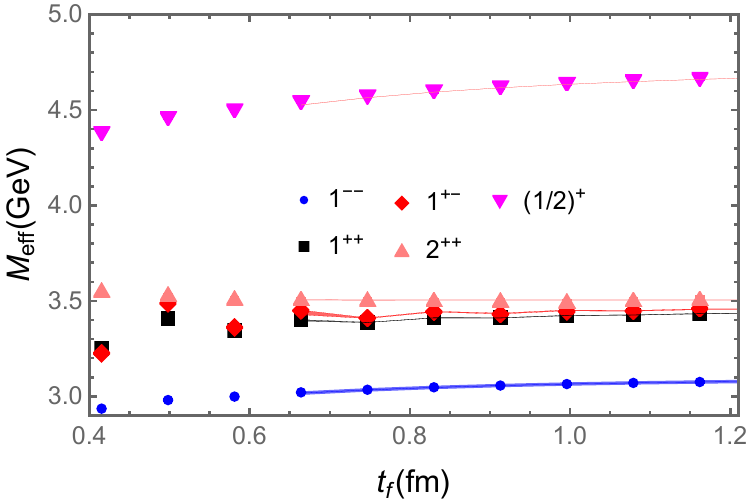} \\ 
	\includegraphics[width=0.4\textwidth]{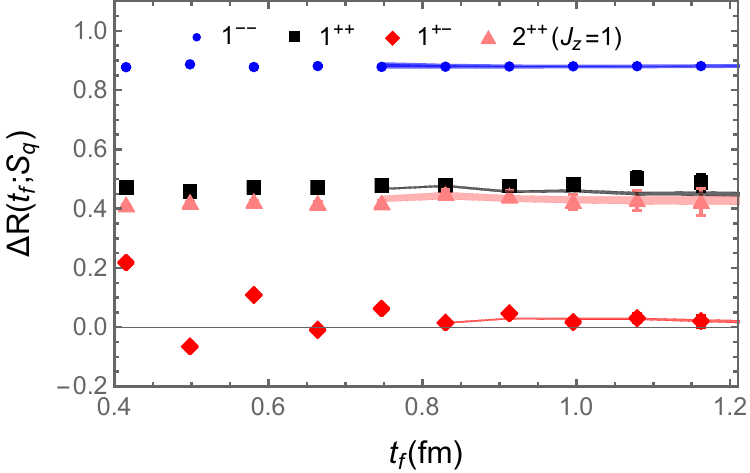} \\ 
    \includegraphics[width=0.4\textwidth]{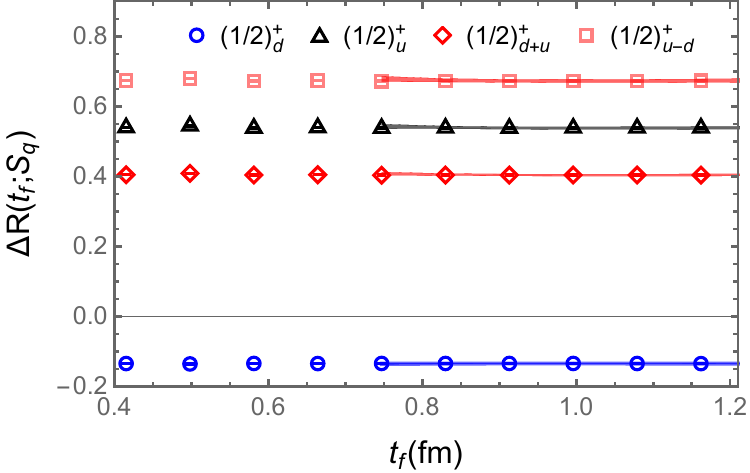}
	\caption{The effective masses (upper panel) and differential ratio $\Delta_R(t_f,S_q)$ for the quark spin operator in the different charmed hadron states. The middle panel shows the results for spin one and spin two meson states. The bottom panel is the results of $(1/2)^+$ state. The bands correspond to the multi-state fits of the results, and the subscripts u and d represent the results of u-type heavy quark and d-type heavy quark, respectively. Note that the renormalization constant $Z_A$ has been multiplied in these results.}
	\label{fig:quarkspin}
\end{figure}

{\it Result:} The effective mass $M_{\rm eff}$, defined as $\mathrm{log}[C(t_f)/C(t_f+a)]/a$, for different charmed hadron states is presented in the upper panel of Fig~\ref{fig:quarkspin}, with curves corresponding to the two-state fit described later in this section. We observe that $M_{\rm eff}$ approaches a stable value in the limit $t_f\rightarrow \infty$, which corresponds to the ground state mass with different quantum numbers.

For the quark spin operator, the renormalization constants of the axial-vector current $Z_A$ has been calculated in Ref.~\cite{He:2022lse}. In the middle panel of Fig.~\ref{fig:quarkspin}, the blue date points correspond to the results of $1^{--}$ channel, where it can be observed that the quark spin contributes a significant portion to the total spin of $1^{--}$ charmonium. Such an observation is comparable to the expectations from the quark model, where the $1^{--}$ channel corresponds to the 1S channel and the entire spin is attributed to the quark spin. The black and orange data represent the results in the $1^{++}$ and $2^{++}(J_z=1)$  channels, respectively. The contributions of quark spin to these two channels' spin are very close since they are corresponding to the same spin-orbital triplet with angular momentum $L=1$ and total quark spin $S=1$, and then both $S$ and $L$ should contribute to the total AM equally. The red data points represent the result of $1^{+-}$ channel which corresponds to $S=0$ and $L=1$ in the quark model, one can see the quark spin has little effect on hadron spin. 
 
In addition to the meson case, we also consider the  contribution of the u-type heavy quark and d-type heavy quark to the spin $(1/2)^+$ triple heavy quark state (the bottom panel of Fig.~\ref{fig:quarkspin}), with the valence quark mass equals to that of charm quark. Their contribution differ by a factor $-4.0(1)$, which is consistent with the prediction of SU(6) quark model $\frac{4}{3}/(-\frac{1}{3})=-4$. The quantity $\Delta u-\Delta d\sim 1.513(38)$ is also close to the quark model prediction 5/3 in Eq.~(\ref{eq:delta_ud}).

\begin{figure}[htbp] 
\includegraphics[width=0.4\textwidth]{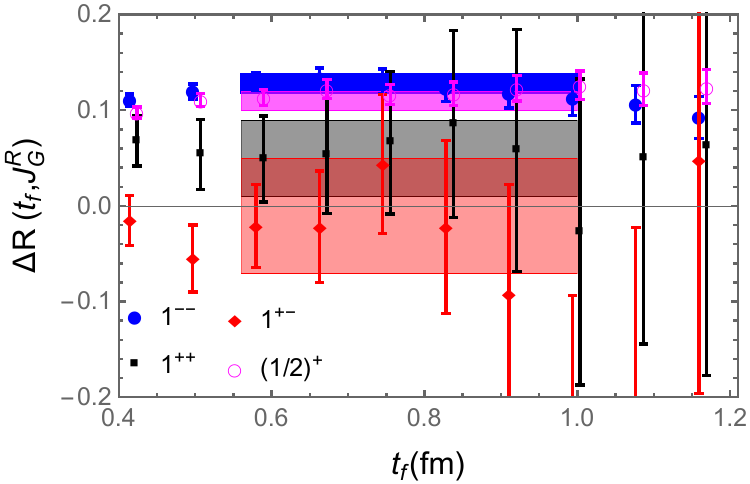} 
\caption{The differential ratio $\Delta_R (t_f, J^R_G)$ of renormalized gluon AM $J^R_G$ in the different charmonium states. The bands are constructed using the fit results. Note that the renormalization constant $Z_{gg}$ has been multiplied in these results.}
\label{fig:final_JG}
\end{figure}

The results of differential ratio $\Delta_R (t_f, J^R_G)$ of renormalized gluon AM $J_G^R=Z_{gg}J_G$ with $Z_{gg}$ from Ref.~\cite{Yang:2018nqn} for these channels are shown in Fig.~\ref{fig:final_JG}. In the appendix, we also investigate the bare matrix elements of gluon spin operator $E\times A$ under Coulomb gauge~\cite{Lorce:2012rr,Zhao:2015kca,Yang:2016plb} in these states, the results are comparable with the gluon AM presented here.

After removing the effect from mixing operators, we can use two state fit to extract the ground state matrix elements. The fit ansatz for the two point correlation functions and quark three correlation functions are~\cite{He:2021bof} 
\bea
\label{eq:fitans}
C_2(t_f;H)&=&B_0 e^{-M_H t_f} (1+B_1 e^{-\delta m t_f}),
 \nonumber\\
C^{i}_3(t_f)&=&e^{-M_Ht_f}\Big(B_0 t_f \langle O^{i}\rangle_H \nonumber\\
 &+& B^{i}_{2}e^{-\delta m t_f} + B^{i}_3t_f e^{-\delta m t_f}+B^{i}_4\Big)
\eea
The effective masses of $1^{++}$ and $1^{+-}$ in Fig.~\ref{fig:quarkspin} show the more obvious oscillation behavior than those of other channels, which is related to the unphysical mode of the domain wall fermions~\cite{Liang:2013eoa}. To describe such a behavior, we add a $(-1)^t$ term to the fit ansatz of 2pt~\cite{Sun:2020pda}, that is 
\bea
C_2(t_f;H)&=&B_0 e^{-M_H t_f} (1+B_1 e^{-\delta m t_f})
\nonumber\\
&+&W(-1)^t e^{- M_2 t_f},
\eea
in the fitting of other channels' data, we still use the fit ansatz defined in Eq.~(\ref{eq:fitans}).
For the $2^{++}$ channel, since we do not have large statistics as spin one case, so we just assume they should be equivalent to the $1^{++}$ because they are belong to the spin-triplet states, and their quark spin components are also consistent within the uncertainties. 

Considering that the statistical uncertainty of the gluon angular momentum is considerably greater then the quark spin, which reduces the significance of excited state contamination, we conduct the constant fit over the range $t_f \in [0.6, 1.0]$ fm. We then determine the systematic uncertainty associated with excited state contamination by observing the changes in the central value when we adjust the fit range to $t_f \in [0.8, 1.2]$ fm. The fit results for $J^R_G$ are presented in Table.~\ref{tab:renor_table}, the uncertainties following the center values represent statistic and systematic errors, respectively.

\begin{figure*}
\begin{center}
\includegraphics[width=1\textwidth]{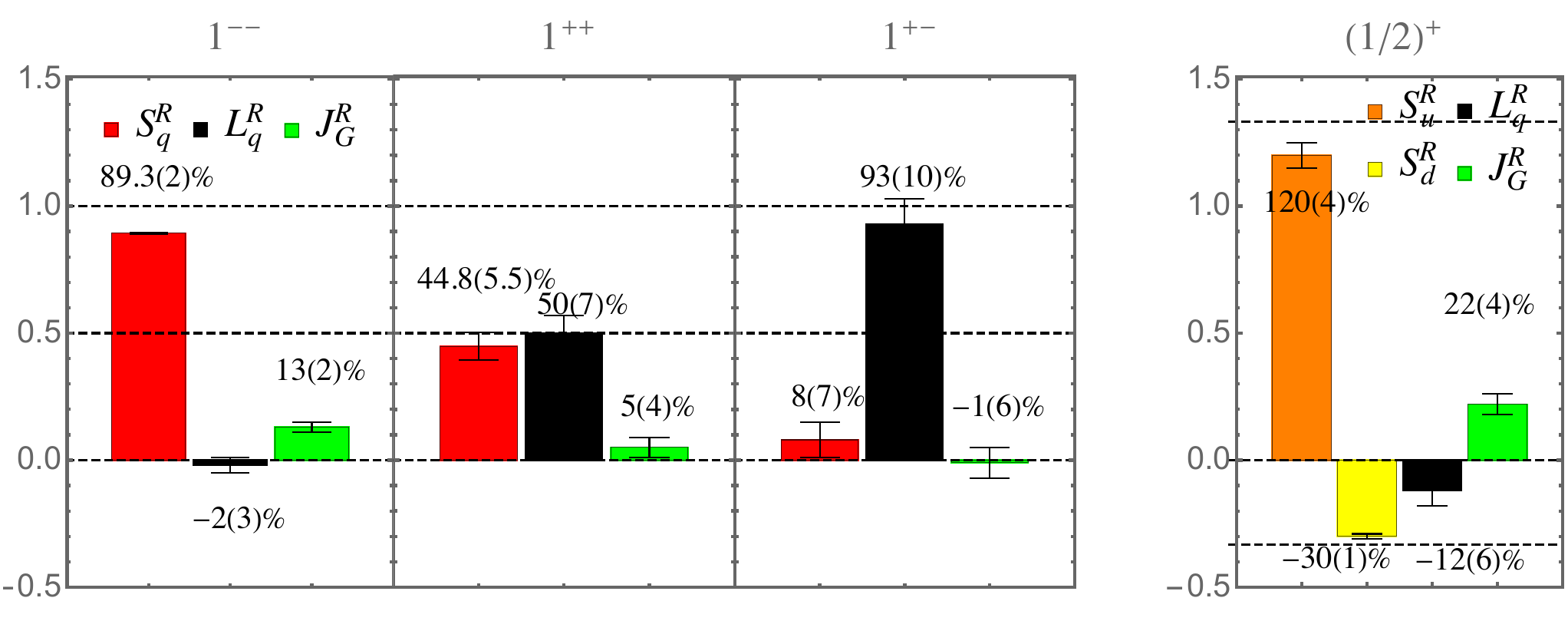}   
\caption{The proportions of quark spin ($S_q^R$), quark orbital angular momentum ($L_q^R$) and gluon total angular momentum ($J_G^R$) in different hadron states.}
\label{fig:piechart}
\end{center}
\end{figure*}

In this calculation, we get the contributions of quark orbital angular momentum $L_q$ by assuming the total contributions from renormalized quark spin ($S_q^R$), quark orbital angular momentum ($L_q^R$) and gluon angular momentum ($J_G^R$) should add up to the spin of hadrons,
\begin{align}
    L_q^R=1-S_q^R-J_G^R.
\end{align}
Note that we ignore the disconnected quark spin contribution and also the mixing between quark and gluon AM, since they should be highly suppressed in the heavy quark system.
The summary of renormalized results are listed in Table.~\ref{tab:renor_table}. The bar charts describing the proportions of quark spin ($S_q^R$), quark orbital angular momentum ($L_q^R$) and gluon total angular momentum ($J_G^R$) in different hadron states are shown in Fig.~\ref{fig:piechart}.

\begin{table*}[t]
	\caption{\label{tab:renor_table} The renormalized results of total quark spin ($S_q^R$), quark orbital angular momentum ($L_q^R$) and gluon total angular momentum ($J_G^R$).}
	\begin{ruledtabular}
		\begin{tabular}{cccc}
			 $J^{PC}$ &   $S_q^R$ & $L_q^R$ & $J_G^R$
			\\\hline
             $(1/2)^+$ &  0.453(24)  &  \!$-0.06(03)$      &   \ \ 0.11(01)(01)  \\
			 $1^{--}$ & 0.893(03) &  \!$-0.02(03)$  &   \ \ 0.13(01)(01) \\
			 $1^{++}$  &  0.448(55)  &\ \ 0.50(07) &   \ \  0.05(04)(01) \\
			 $1^{+-}$   &   0.080(70)    &\ \ 0.93(10)   &  \!$-0.01(06)(04)$ \\
			 $2^{++}(J_z=1)$ &  0.436(11)  &\ \ 0.50(07)     &   \ \ 0.05(04)(01) \\
    
		\end{tabular}
	\end{ruledtabular}
\end{table*}

{\it Summary:} In this study, we investigate the spin decomposition of 1S, 1P and $(1/2)^{+}$ charmonium states. The contributions of the quark spin to the spin of these charmonium states are comparable with the expectations of quark model. For instance, the quark spin contributes about 90\%\ to the total spin of $1^{--}$, which is consistent with the expectation that this channel corresponds to an S-wave state where the spin is entirely contributed by the quark spin in the quark model. For the $1^{++}$ and $2^{++}$ channels, their decomposition results are very similar, these states belong to the spin triplet with angular momentum $L=1$ and quark spin $S=1$. We also find the contribution of quark spin is consistent with zero in the $1^{+-}$ state, and such a state is the P wave singlet state in the quark model. For the $(1/2)^+$ state, the ratio $\Delta u/\Delta d=-4.0(1)$ agrees with the quark model prediction perfectly, and the values $\Delta u=1.20(4)$ and $\Delta d=-0.30(1)$ are 90\% of the predicted values.

On the other hand, the calculation~\cite{Liang:2018pis} on the same ensemble, extrapolated to the physical light quark mass, predicts $\Delta u=0.85(4)(3)$ and $\Delta d=-0.41(2)(2)$, which are consistent with the experimental values, with $\Delta u/\Delta d=-2.1(2)(1)$. Such an observation indicates that the complexity of proton and light hadron spin decomposition primarily originates from the relativistic effects of the light quark which suppress the quark spin contribution significantly. 

Conversely, the substantial gluon angular momentum contribution in the spin $(1/2)^+$ state with triple heavy quarks with the charm quark mass, remains sizable at $\sim20$\% compared to that with the physical light quark (40\%-50\%~\cite{Alexandrou:2017oeh,Wang:2021vqy}), highlighting the ongoing importance of the gluon in the realm of charm  baryon physics. Further study using different quark masses should be helpful to provide a quantitative criteria on the effectiveness of quark model as a function of quark mass.

We also obtain the contribution of gluon total angular momentum through calculating the EMT matrix elements in different charmonium states. The contribution of gluon is small among the hadron states considered in this work, and the results of $1^+$  channels are very noisy. The more precise calculations on gluon contribution, among with the mixing effects between quark and gluon AM are deserved further investigation in the future.

\section{Acknowledgement}
We thank Heng-Tong Ding, Xiangdong Ji, Luchang Jin, Keh-Fei Liu, Jianhui Zhang, and Jian Zhou for valuable discussions, and the RBC and UKQCD collaborations for providing us their DWF gauge configurations. The calculations were performed using the GWU-code~\cite{Alexandru:2011ee,Alexandru:2011sc} through HIP programming model~\cite{Bi:2020wpt}. The numerical calculation were carried out on the ORISE Supercomputer, HPC Cluster of ITP-CAS and Advanced Computing East China Sub-center. F.H. is supported by the National Science Foundation under CAREER Award PHY-1847893 and the LANL Laboratory Directed Research and Development (LDRD) program. This work is supported in part by NSFC grants No. 12293060, 12293062, 12175073, 12222503 and 12047503, the Natural Science Foundation of Basic and Applied Basic Research of Guangdong Province under Grant No.\ 2023A1515012712, the Strategic Priority Research Program of Chinese Academy of Sciences, Grant No.\ XDB34030303 and YSBR-101, the science and education integration young faculty project of University of Chinese Academy of Sciences, and also a NSFC-DFG joint grant under Grant No.\ 12061131006 and SCHA 458/22.

\clearpage

\appendix

\section{Extract gluon AM of different spin one meson from form factors}

The GFFs of spin one particle can be expressed 
as~\cite{Polyakov:2019lbq}
\begin{widetext}
\bea\label{Eq:EMT-FFs-spin-1}
\langle p^\prime,\sigma^\prime| \hat T_{\mu\nu}^a(0) |p,\sigma\rangle
&=& \biggl[
2\bar{P}_\mu \bar{P}_\nu  \Bigl(
- {\epsilon^{\prime*}\cdot \epsilon} \, A^a_0 (q^2)
 +{ {\epsilon^{\prime*}\cdot \bar{P}} \, {\epsilon \cdot \bar{P}} \over m^2}
 \, A^a_1(q^2) \Bigl)
\nonumber\\
&&+2\left[\bar{P}_\mu(\epsilon^{\prime*}_{\nu} \,\epsilon\cdot \bar{P}+\epsilon{\nu}\,
\epsilon^{\prime*}\cdot \bar{P})
+\bar{P}_\nu(\epsilon^{\prime*}_{\mu}\, \epsilon\cdot
\bar{P}+\epsilon_{\mu} \,\epsilon^{\prime*}\cdot
\bar{P}) \right] \, J^a (q^2)
\nonumber\\
&&+\frac12(q_\mu q_\nu-g_{\mu\nu}q^2)
 \Bigl(
{\epsilon^{\prime*}\cdot \epsilon} \, D^a_0 (q^2)
+{ {\epsilon^{\prime*}\cdot \bar{P}} \, {\epsilon \cdot \bar{P}} \over m^2}  \, D^a_1(q^2)\Bigl)
\nonumber\\
&&+\Bigl[\frac12(\epsilon_{\mu}
\epsilon^{\prime*}_{\nu}+\epsilon^{\prime*}_{\mu}\epsilon_{\nu})q^2
-(\epsilon^{\prime*}_{\mu}q_\nu+\epsilon^{\prime*}_{\nu} q_\mu)\,\epsilon\cdot \bar{P}
\nonumber\\
&&
+(\epsilon_{\mu} q_\nu+\epsilon_{\nu}
q_\mu)\,\epsilon^{\prime*}\cdot \bar{P}
-4g_{\mu\nu} \, {\epsilon^{\prime*}\cdot \bar{P}} \, {\epsilon\cdot \bar{P}} \Bigl] \, E^a(q^2)
\nonumber\\
&&
+\Bigl(\epsilon_{\mu}
\epsilon^{\prime*}_{\nu}+\epsilon^{\prime*}_{\mu}\epsilon_{\nu} - \frac{{\epsilon^{\prime*}\cdot \epsilon}}{2}\,g_{\mu\nu} \Bigl)
\,{m^2} \, {\bar f}^a (q^2) \nonumber\\
&&+g_{\mu\nu} \Bigl( {\epsilon^{\prime*}\cdot \epsilon} \, {m^2}\, {\bar c}^a_0(q^2)\, +  \, {\epsilon^{\prime*}\cdot \bar{P}} \, {\epsilon \cdot \bar{P}}
\,{\bar c}^a_1(q^2)  \Bigl)  \biggr]  \ ,
\eea
\end{widetext}
with the normalization of one-particle states is expressed as $\langle p^\prime,\sigma^\prime |p,\sigma\rangle=2p^0\delta^3(\vec{p'}-\vec{p})(2\pi)^3$.

Based on those form factors, the gluon AM of the $J=1$ meson can be expressed as~\cite{Polyakov:2019lbq}
\begin{align}
    J^V_G=[J^g(q^2)+\frac12\bar{f}^g(q^2)]_{q^2\rightarrow 0}.\label{eq:meson_AM_sm}
\end{align}

As we have noted in the main text, the GFFs $J^g(q^2)$ and $\bar{f}^g(q^2)$ can be extracted from the following equations, 
\begin{subequations}\label{eq:EMgluon_sm}
\begin{align}\label{eq:EMgluon1_sm}
&\Delta R^g(y,z,T^g_{4y},\vec{q})  \nonumber\\
&=-{\rm i}\frac{k}{4E_f}\left[(E_f+m)J^g(q^2)-(E_f-m)E^g(q^2)\right], \\
\label{eq:EMgluon2_sm}
&\Delta R^g(z,y,T^g_{4y},\vec{q}) \nonumber\\
&={\rm i}\frac{k}{4E_f}\Big[\frac{(E_f+m)(2m^2+k^2)}{2m^2}J^g(q^2)\nonumber\\
&\quad+\Big((E_f-m)-\frac{k^2(E_f+m)}{2m^2}\Big)E^g(q^2)+2E_f\bar{f}^g(q^2)\Big], \\
\label{eq:EMgluon3_sm}
&\Delta R^g(x,y,T^g_{xy},\vec{q})=\left[-\frac{k^2}{2}E^g(q^2)+m^2\bar{f}^g(q^2)\right]\frac{1}{2E_f},
\end{align}
\end{subequations}
where $q^2=(E_f-m)^2-k^2=-k^2(1+{\cal O}(\frac{k^2}{m^2}))$. 

\begin{figure}[htbp] 
\includegraphics[width=0.4\textwidth]{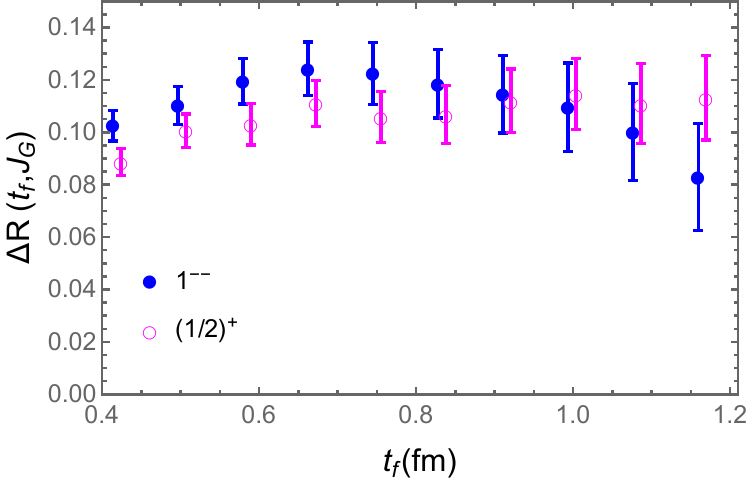}  \\ 
\includegraphics[width=0.4\textwidth]{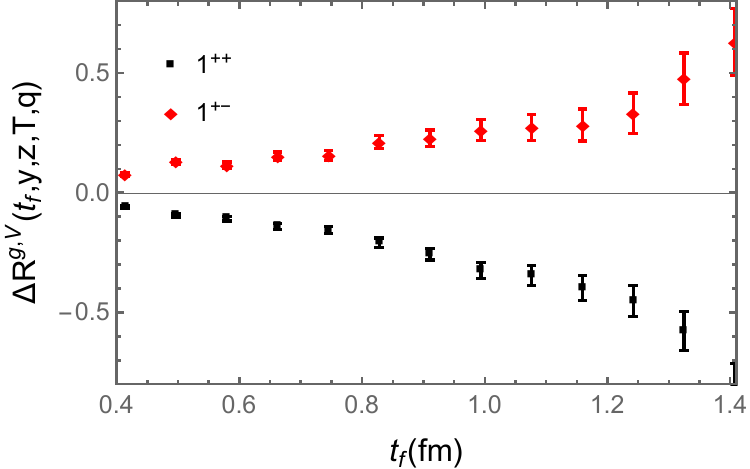} 
\\[-0.4cm]
\caption{The differential ratio $\Delta_R^g (t_f)$ for the different gluon operator in the different charmonium states. The upper panel is the result of total gluon angular momentum operator in the $1^{--}$ (blue dots) and $(1/2)^+$ (purple circles) states. One can observe a plateau when $t_f$ is large. In the other hand, $\Delta R^{g,V}(t_f,z,y,T^{4y},\vec{q})$ (black data points, for $1^{++}$) and $\Delta R^{g,V}(t_f,y,z,T^{4y},\vec{q})$ (red data points, for $1^{+-}$) shown in the bottom panel are increasing when $t_f$ is increasing, due to the operator mixing.}
\label{fig:ori_JG}
\end{figure}

After obtaining the three point correlation functions with the combination of different components of gluon energy momentum tensor and hadrons' polarizations as listed in Eq.~(\ref{eq:EMgluon_sm}), one can solve these equations to get the differential ratio of gluon total angular momentum versus $t_f$ for vector meson. The differential ratio for the $(1/2)^+$ state is calculated using $T^{44}$ component. The results of differential ratio for $(1/2)^+$ and $1^{--}$ channel are plotted in the upper panel of Fig.~\ref{fig:ori_JG}, a clear plateau can be observed when $t_f$ is large. 

However, this approach is not applicable to the $1^{++}$ or $1^{+-}$ channels due to the operator mixing when we boost the final state to the moving frame. As illustrated in the bottom panel of Fig.~\ref{fig:ori_JG}, for certain form factor combinations of the $1^{+-}$ channel (represented by $\Delta R^{g}(y,z,T^{4y},\vec{q})$ defined in Eq.~(\ref{eq:EMgluon1_sm}), shown as red data points) and the $1^{++}$ channel (represented by $\Delta R^{g}(z,y,T^{4y},\vec{q})$ defined in Eq.~(\ref{eq:EMgluon2_sm}), shown as black data points), the values do not saturate at the large $t_f$ limit, due to the mixing with lighter hadron. Thus it is impossible to solve the needed form factors using Eqs.~(\ref{eq:EMgluon1_sm}-\ref{eq:EMgluon3_sm}).

\begin{figure}[htbp] 
\includegraphics[width=0.4\textwidth]{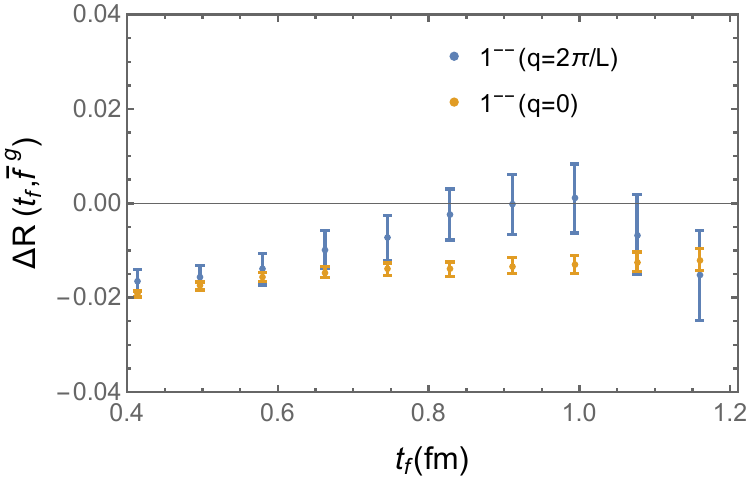}  \\ 
\includegraphics[width=0.4\textwidth]{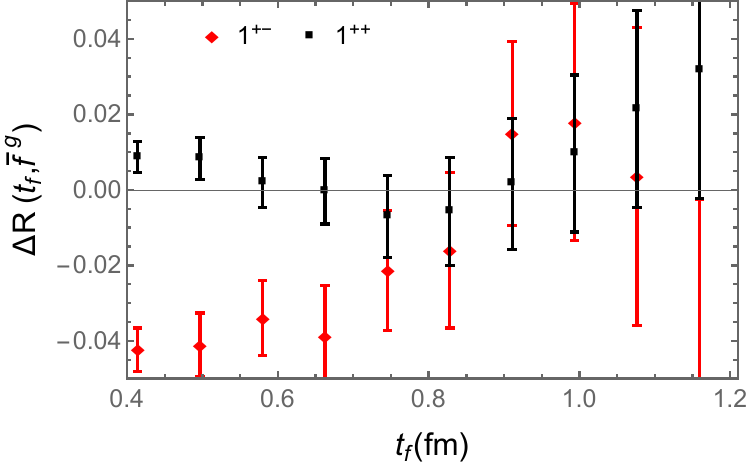} 
\caption{The differential ratio for the form factor $\bar{f}^g$ in the different charmonium states. The upper panel is the result of form factor $\bar{f}_g$ at the rest frame and smallest momentum $\vec{q}=(0,0,2\pi/L$), the result at the smallest momentum is solved using Eq.~(\ref{eq:EMgluon1_sm}),~(\ref{eq:EMgluon2_sm}) and ~(\ref{eq:EMgluon3_sm}). The lower panel shows the results of form factor $\bar{f}_g$ for $1^{++}$ and $1^{+-}$ channel at the rest frame.}
\label{fig:fbar}
\end{figure}

\begin{figure}[htbp] 
\includegraphics[width=0.4\textwidth]{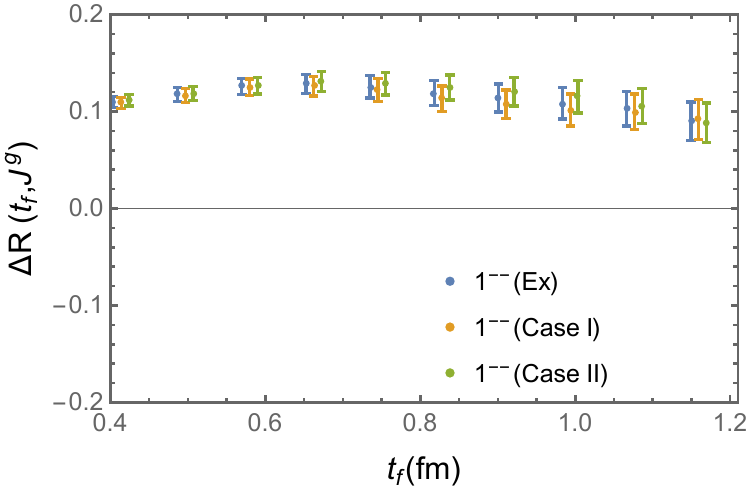}  \\ 
\includegraphics[width=0.4\textwidth]{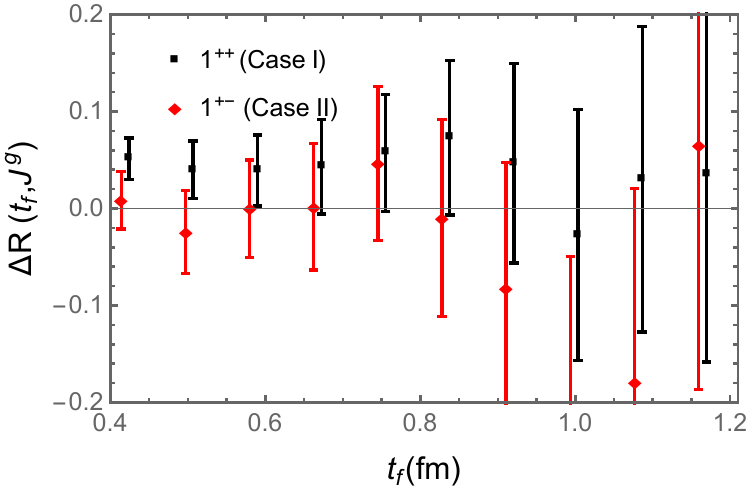} 
\caption{The differential ratio for form factor $J^g(t)$ in the different charmonium states. The upper panel is the result of form factor $J^g$. ``Ex" represent the results obtained by solving Eq.~(\ref{eq:EMgluon1_sm}),~(\ref{eq:EMgluon2_sm}) and ~(\ref{eq:EMgluon3_sm}).}
\label{fig:final_Jg}
\end{figure}

Fortunately, the other two equations in Eq.~(\ref{eq:EMgluon_sm}) are free from this mixing, allowing one to approximate $\bar{f}^g(t)$ by $\bar{f}^g(0)$ and solve for $J^g$ from them. The upper panel of Fig.~\ref{fig:fbar} displays the differential ratio for $\bar{f}_g$ in the rest frame (yellow data points) and at the smallest non-zero momentum (blue data points), showing consistency within two sigma. The lower panel of the same figure presents the $\bar{f}_g$ results of $1^{+-}$ (red data points) and $1^{++}$ (black data points) in the rest frame, indicating that the values in all cases studied are small. Consequently, their contribution to the total meson spin will be at the 1\% level, considering the factor 1/2 in the definition Eq.~(\ref{eq:meson_AM_sm}).

Then one can solve the form factor $E^g(t)$ by using Eq.~(\ref{eq:EMgluon3_sm}) and assuming $f^g(q^2)\approx f^g(0)$,
\bea
E^g(q^2)=-\frac{2}{k^2}\Delta R^g(x,y,T^{xy},\vec{q})-m^2\bar f^g(0),
\eea
and use Eq.~(\ref{eq:EMgluon1_sm}) to obtain the form factor $J^a(t)$ for the $1^{++}$ channel, 
\bea
&&J^g(q^2)=\frac{4{\rm i}E_f\Delta R^g(y,z,T^{4y},\vec{q})}{(m+E)k}+\frac{E-m}{m+E}E^g(q^2),\nonumber\\
\eea
and we define refer to it as `Case I'. 

Analogous, by using Eq.~(\ref{eq:EMgluon2_sm}), we can obtain the results of $J^{g}(q^2)$ for the $1^{+-}$ channel, which we denote it as `Case II',
\bea
J^g(q^2)&=&-\frac{8 i\Delta R^g(z,y,T^{4y},\vec{q}) E_f m^2}{k (E_f+m) \left(k^2+2 m^2\right)}
\nonumber\\
&-&\frac{4 E_f m^2}{(E_f+m) \left(k^2+2 m^2\right)}\bar{f}^g(q^2)  \nonumber\\
&+&\frac{\left(E_f k^2-2 E_f m^2+k^2 m+2 m^3\right)}{(E_f+m) \left(k^2+2 m^2\right)}E^g(q^2) , \nonumber\\
\eea

The results of differential ratio for the gluon GFF $J^g$ in the $1^{--}$, $1^{++}$ and $1^{+-}$ channel are shown in Fig.~\ref{fig:final_Jg}. In the upper panel, we present a comparison of the results of $1^{--}$ obtained using all three equations (labelled by `Ex') with those obtained using the method defined in `Case I' and `Case II'. It can be observed that the results obtained using three different strategies exhibit consistency within the uncertainties. The results of the $1^{++}$ and $1^{+-}$ channel are shown in the bottom panel, those results do not suffer from the operator mixing problem and they are consistent with zero within the uncertainties. Note that $J^g$ is roughly $-5$ times of $f^g$ in the $1^{--}$ case and dominates in the bare $J/\psi$ glue AM.

\section{Bare matrix element of gluon spin}
\begin{figure}[htbp] 
\includegraphics[width=0.4\textwidth]{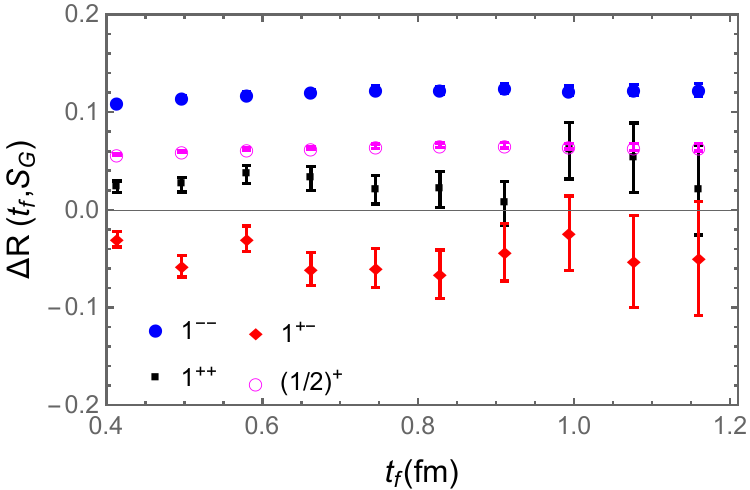} 
\caption{The results of gluon spin in different hadron states.}
\label{fig:DeltaR_SG}
\end{figure}

The calculation of gluon spin is similar as that of quark spin, the corresponding operator can be written as~\cite{Lorce:2012rr,Zhao:2015kca,Yang:2016plb}
\begin{align}
\vec{S}_G=2 \int d^3x \text{Tr}(\vec{E}_c\times\vec{A}_c) \end{align}
under Coulomb gauge fixing condition $\vec{\partial}\cdot\vec{A}=0$.  Similar as the calculation of gluon AM, we perform the one-step hyp smearing on the gluon operator, and the differential ratio of gluon spin $\Delta R^g(t_f,S_G)$ is shown in Fig.~\ref{fig:DeltaR_SG}. The bare quantities of gluon spin for $1^{--}$ is very close to that of gluon AM shown in Fig.~\ref{fig:final_JG}. The results of gluon spin tend to be positive in $1^{++}$ channel and negative in $1^{+-}$ channel. For the $(1/2)^+$ case, the bare matrix element of gluon spin is about half of that of gluon AM.

\bibliographystyle{apsrev4-1}
\bibliography{reference.bib}

\end{document}